\documentclass[journal=jpccck,manuscript=article]{achemso}
\SectionNumbersOn 

\usepackage[version=3]{mhchem} % Formula subscripts using \ce{}
\usepackage[utf8]{inputenc}
\usepackage[T1]{fontenc}

\author{Bj{\"o}rn Kirchhoff}
\affiliation[HI]
{Science Institute and Faculty of Physical Sciences, University of Iceland, VR-III, Hjarðarhagi 2, 107 Reykjav\'{\i}k, Iceland.}
\alsoaffiliation[UUlm]
{Institute of Electrochemistry, Ulm University, Albert-Einstein-Allee 47, 89081 Ulm, Germany.}
\author{Christoph Jung}
\affiliation[HIU]{Helmholtz-Institute Ulm (HIU) Electrochemical Energy Storage, Helmholtz-Straße 16, 89081 Ulm, Germany.}
\alsoaffiliation[KIT]
{Karlsruhe Institute of Technology (KIT), P.O. Box 3640, 76021 Karlsruhe, Germany.}
\author{Hannes J{\'o}nsson}
\affiliation[HI]
{Science Institute and Faculty of Physical Sciences, University of Iceland, VR-III, Hjarðarhagi 2, 107 Reykjav\'{\i}k, Iceland.}
\author{Donato Fantauzzi}
\affiliation[HI]
{Science Institute and Faculty of Physical Sciences, University of Iceland, VR-III, Hjarðarhagi 2, 107 Reykjav\'{\i}k, Iceland.}
\author{Timo Jacob}
\email{timo.jacob@uni-ulm.de}
\affiliation[UUlm]
{Institute of Electrochemistry, Ulm University, Albert-Einstein-Allee 47, 89081 Ulm, Germany.}
\alsoaffiliation[HIU]
{Helmholtz-Institute Ulm (HIU) Electrochemical Energy Storage, Helmholtz-Straße 16, 89081 Ulm, Germany.}
\alsoaffiliation[KIT]
{Karlsruhe Institute of Technology (KIT), P.O. Box 3640, 76021 Karlsruhe, Germany.}

\title[]{Simulations of the Electrochemical Oxidation of Pt Nanoparticles of Various Shapes}

\abbreviations{}

\begin{document}

\begin{tocentry}

%Some journals require a graphical entry for the Table of Contents.
%This should be laid out ``print ready'' so that the sizing of the

%Inside the \texttt{tocentry} environment, the font used is Helvetica
%8\,pt, as required by \emph{Journal of the American Chemical

%The surrounding frame is 9\,cm by 3.5\,cm, which is the maximum
%permitted for  \emph{Journal of the American Chemical Society}
%graphical table of content entries. The box will not resize if the
%content is too big: instead it will overflow the edge of the box.

%This box and the associated title will always be printed on a
%separate page at the end of the document.

\includegraphics{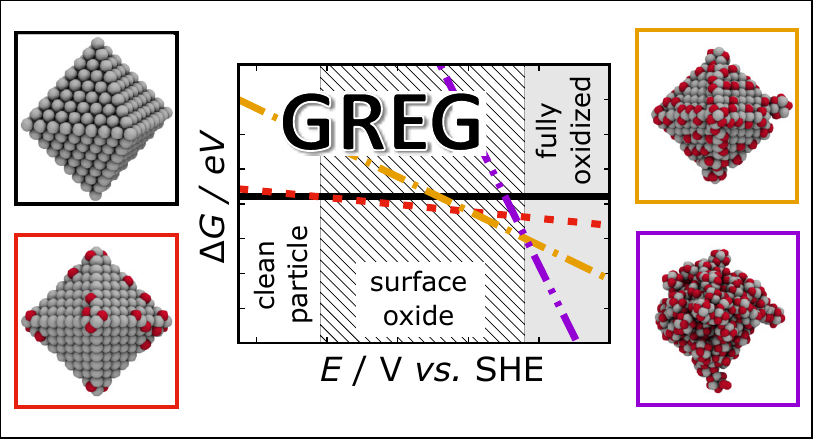}

\end{tocentry}

\begin{abstract}
  The activity and stability of a platinum nanoparticle (NP) is not only affected by its size but additionally depends on its shape. To this end, simulations can identify structure-property relationships to make \textit{a priori} decisions on the most promising structures. While activity is routinely probed by electronic structure calculations on simplified surface models, modeling the stability of NP model systems in electrochemical reactions is challenging due to the long timescale of relevant processes such as oxidation beyond the point of reversibility. In this work, a routine for simulating electrocatalyst stability is presented. The procedure is referred to as GREG after its main ingredients - a \textit{grand-canonical} simulation approach using \textit{reactive} force fields to model \textit{electrochemical} reactions as a function of the \textit{galvanic} cell potential. The GREG routine is applied to study the oxidation of 3~nm octahedral, cubic, dodecahedral, cuboctahedral, spherical, and tetrahexahedral platinum NPs. The oxidation process is analyzed using adsorption isobars as well as interaction energy heat maps that provide the basis for constructing electrochemical phase diagrams. Onset potentials for surface oxidation increase in the sequence cube $\approx$ dodecahedron $\leq$ octahedron $\leq$ tetrahexahdron $<$ sphere $<$ cuboctahedron, establishing a relationship between oxidation behavior and surface facet structure. The electrochemical results are rationalized using structural and electronic analysis.
\end{abstract}

\section{\label{sec:introduction}Introduction}
Platinum nanoparticles (NPs) are relevant for a wide range of electrocatalytic processes, including small molecule oxidation,\cite{sun2019} plasma catalysis,\cite{puliyalil2018} the oxygen evolution reaction (OER),\cite{suen2017} and the oxygen reduction reaction (ORR).\cite{sui2017} Due to the scarcity and price of Pt, development of material optimization strategies aimed at minimizing Pt loading and maximizing mass activity is a critical task. To this end, shape-controlled synthesis is being discussed as a powerful concept to tune activity and stability of Pt NP electrocatalysts.

The relationship between shape and catalytic activity of Pt NPs has been studied extensively; %For example, Narayanan \textit{et al.},\cite{narayanan2004} Wang \textit{et al.},\cite{wang2008} and Sánchez-Sánchez \textit{et al.}\cite{sanchez-sanchez2010} measured the electrochemical ORR activity of various particle shapes - including tetrahedral, cubic, octahedral, and spherical NPs - noting pronounced dependence between the dominant facet and the resulting catalytic properties. Cui \textit{et al.} measured the ORR activity of octahedral \ce{PtNi} alloy NPs using aberration-corrected scanning transmission electron microscopy and electron energy-loss spectroscopy.\cite{cui2013} They observed significantly higher ORR activity compared to a non-shape-controlled commercial Pt reference catalyst. Tian and co-workers carried out electrochemical synthesis of large tetrahexahedral NPs, covered in \{730\} surface facets, and measured good catalytic activity for the electro-oxidation of small organic molecules.\cite{tian2007,tian2008} 
reviews by Kleijn \textit{et al.},\cite{kleijn2014} Sui \textit{et al.},\cite{sui2017} and Strasser \textit{et al.}\cite{strasser2018} summarize experimental findings on this topic. The present work focuses on the relationship between particle shape and stability. In many catalytic applications such as the OER, ORR, small molecule oxidation, or plasma catalysis, Pt NP electrocatalysts are in contact with elemental oxygen. As a result, oxidative degradation has been identified as a key limiting factor for electrocatalyst longevity.\cite{zhang2009,hidai2012,meier2014} While computational methods are frequently used to probe the catalytic activity of NPs, usually by means of simplified surface model systems, degradation processes are rarely studied in simulations for several reasons.

Firstly, computational treatment of shaped NPs is methodologically challenging. Typical Pt NPs used in applications are 2--5~nm in diameter and thus contain too many atoms (250--4000 atoms) for treatment with electronic structure methods such as Density Functional Theory (DFT). Therefore, DFT calculations of high-indexed surface models have been used to approximate the properties of NPs. Examples include the works of Zhu \textit{et al.}, where the potential- and oxidation-induced formation of tetrahexahedral NPs was calculated,\cite{zhu2013} and Tritsaris \textit{et al.}, who used DFT adsorption energy results from a \{544\}-indexed surface as basis for ORR activity estimation of cuboctahedral NPs.\cite{tritsaris2011} However, Calle-Vallejo and co-workers discuss that high-indexed surface models cannot sufficiently represent all properties of NPs; as an example, they note that stepped surface models contain both convex and concave surface sites while typical NPs only contain one type.\cite{calle-vallejo_why_2017}

A natural choice for simulating NPs, therefore, are potential functions rooted in classical mechanics, which can be used to routinely calculate 10,000s of atoms. However, classical force field simulations require fixed bonding tables, leading to mostly rigid systems that cannot undergo chemical reactions during simulation time. This limitation hampers their applicability to simulations of degradation processes such as deep oxidation, \textit{i.e.} beyond straightforward surface adsorption. Examples of classical force fields applied to NP models include studies by Wen \textit{et al.}, who calculated the melting temperature of NPs using classical dynamics,\cite{wen2009,wen2009_2} and Huang \textit{et al.}, who computed the stability of various NP shapes as a function of the particle size.\cite{huang2011} Both groups used Sutton-Chen-type potential functions. Fortunately, the gap between electronic structure calculations and molecular mechanics methods can be bridged. The ReaxFF\cite{vanduin2001} reactive force field approach overcomes the limitations of potentials rooted in classical mechanics by introducing bond order dependency into the potential terms, thereby allowing bonds to form or break during simulation time. 

While the ReaxFF method in principle allows us to study (electro-)chemical reactions on NP catalysts, a final piece is missing to enable efficient simulations of electrochemical processes: because the large size of the model system prohibits manual generation of input structures, an automated method is required to efficiently sample the configuration space spanned by the initial NP model system, the reactants, and the electrochemical conditions (temperature, pressure, pH, galvanic cell potential, \textit{etc.}). One possible solution is to combine ReaxFF with a grand-canonical Monte Carlo algorithm.\cite{senftle2013,senftle2014} %Fantauzzi \textit{et al}. used a Pt/O reactive force field\cite{fantauzzi2014} in combination with a grand-canonical Monte Carlo (GCMC) algorithm\cite{senftle2013,senftle2014} to predict stable oxide surface structures on a large platinum \{111\} model surface.\cite{fantauzzi_growth_2017} The predicted structures have since been experimentally confirmed using X-ray photon spectroscopy (XPS) at near-ambient pressure (NAP) conditions ($p_\text{O} \approx 1$~mbar).
In the ReaxFF-GCMC scheme,\cite{senftle2013} reactants are randomly inserted, moved, or deleted in a simulation box until thermodynamic equilibrium with a virtual reservoir is reached. After each random move, the energy is minimized with respect to the atomic coordinates. Because ReaxFF allows for bond formation and breaking, the GCMC approach therefore enables coarse-grained studies of longer time scale processes such as complete oxidation. To the best of our knowledge, only two studies exploited the ReaxFF-GCMC approach to study Pt NP oxidation so far; Gai \textit{et al.} calculated oxygen adsorption isotherms for octahedral, cubic, and cuboctahedral particles\cite{gai2016} and our group simulated the oxidation of cuboctahedral NPs of 2-4~nm size.\cite{kirchhoff2019} Besides the GCMC scheme, the ReaxFF reactive molecular dynamics (MD) approach constitutes another powerful tool to investigate oxidation processes in the time frame of ps to ns. Notable works in this regard include those by Zhu and co-workers who used a ReaxFF-MD approach to study the oxidation of large, stepped Cu surfaces used as model systems for supported, island-like Cu cluster.\cite{Zhu2015,Zhu2016,Zhu2017}

In the present work, we present a simulation routine which combines the ReaxFF-GCMC approach with extended \textit{ab initio} thermodynamics (EAITD) theory\cite{jacob2007} to study catalyst oxidation under electrochemical conditions. The routine is referred to as GREG for brevity; the acronym reflects the key ingredients to this strategy: it uses a \textit{grand-canonical} simulation approach and \textit{reactive} force fields to determine the stability of a model system undergoing an \textit{electrochemical} reaction as a function of the \textit{galvanic} cell potential. Based on a given model system and reactant, the GREG routine can be used to predict the thermodynamically most favorable structure at a given electrode potential. The output structures obtained this way can then be used as a refined input model to study, for example, their catalytic activity under these electrochemical conditions.

The GREG routine is applied to shaped Pt NPs in the context of the ORR to investigate the oxidative degradation and \textit{in situ} structure of fuel cell catalysts. The oxidation of 3~nm octahedral, cubic, dodecahedral, cuboctahedral, spherical, and tetrahexahedral NPs is simulated at various oxygen partial pressure and temperature conditions. EAITD theory is used to construct electrochemical phase diagrams from the datasets of oxidized structures to quantify the influence of particle shape on the onset potentials of different stages of oxidation. The obtained trends are rationalized using structural (generalized coordination numbers) and electronic (partial charge distributions, interaction energy heat maps) analysis.

% -----------------------------------------------------------

\section{\label{sec:methodology}Methodology}

\subsection{The combined ReaxFF / grand-canonical Monte Carlo (GCMC) method}
The ReaxFF-GCMC method was originally implemented by Senftle \textit{et al}.\cite{senftle2013, senftle2014} The algorithm randomly inserts, deletes, or moves a reactant $X$ in the simulation box filled with an initial structure $Y$ with constant number of atoms ($\mu_X N_YVT$ ensemble, where $\mu_X(T,p)$ is the chemical potential of the $X$ reservoir). In this work, $X$ is oxygen to study oxidation. The structure generated this way is relaxed with regards to the atomic coordinates using a ReaxFF potential function, and is accepted or dismissed based on an energy criterion depending on $\mu_X(T,p)$. Because ReaxFF uses a bond order dependent representation of the energy terms, bonds can form or break during relaxation cycles. Formally, this process is repeated until equilibrium with the reservoir is reached. However, the GCMC algorithm also accepts MC steps which slightly exceed the $\mu_X(T,p)$-dependent energy criterion in order to enable the simulation to overcome local energy minima.\cite{senftle2013, scm_reaxff} Post-analysis of the generated structures is therefore required to find the global minimum energy structure of a model system under given simulation conditions. The GCMC algorithm operates on a purely thermodynamic basis and provides no insights into kinetic barriers that might be associated with formation of the generated structures.

\subsection{Extended \textit{Ab Initio} Thermodynamics (EAITD)}
The potential-dependent formation energy $\Delta E_\text{F}^\text{\,system} (\Delta \phi)$ of an oxidized NP is calculated as\cite{jacob2007}
\begin{equation}
\Delta E_\text{F}^\text{\,system} (\Delta \phi) = E^\text{\,system} - E^\text{\,ref} - N_\text{O}\ \mu_\text{O} + 2e\ \Delta \phi, \label{eq:potentialdependent}
\end{equation}
where $E^\text{\,system}$ is the total ReaxFF energy of the oxidized NP as obtained from GCMC calculations, $E^\text{\,ref}$ is the total free energy of the oxygen-free reference NP, $N_\text{O}$ is the number of oxygen atoms in the oxidized NP, $\mu_\text{O}$ is the chemical potential of oxygen (where $\mu_\text{O} = \mu_\text{\ce{H2O}} - \mu_\text{\ce{H2}}$), and $e$ is the elementary charge. The last term of eq. \eqref{eq:potentialdependent} implies all oxygen that is consumed in the electrochemical oxidation reaction originates from the water splitting reaction. Standard hydrogen electrode (SHE) conditions corresponding to $T = 298$~K, $p_\text{\ce{H2}} = 1$~bar, and pH~0 are assumed.

\subsection{Solvent-Accessible Surface Area (SASA) and heat maps}
The interaction energy hypersurface (heat map) of atomic oxygen with a NP is estimated via the NP's solvent-accessible surface area (SASA).\cite{Richmond1984} Firstly, the SASA is obtained by distributing a set of 32,000 points along the van der Waals surface of the NP. At each point, the interaction of an oxygen atom to the particle is evaluated by ReaxFF single point calculations, whereby the corresponding probe atom is placed at 0.35 \AA\ from the NP's van der Waals surface. Each of these points is assigned to a platinum atom and all points corresponding to the same atom are averaged for evaluation. Further details can be found in Ref. [23].

\subsection{The GREG routine}

The GREG routine includes the following steps:
\begin{enumerate}
    \item ReaxFF-GCMC simulations are carried out for a chosen combination of model system and reactant at different temperature and reactant partial pressure conditions to ensure appropriate sampling of the relevant configurational space. Individual simulation runs are analyzed (\textit{e.g.}, convergence tests, adsorption isobars, \textit{etc.})
    \item All individual simulation runs of the same model system at different conditions are merged into one dataset for meta analysis. The combined dataset is analyzed using the EIATD approach, using an electrochemical reaction involving the reactant as basis, to obtain adsorption energy results that depend on the standard hydrogen electrode (SHE) potential.
    \item The most stable reaction product structures are identified for increasing reactant coverage densities and are shown as a function of the electrode potential. The most stable structure at a given electrode potential of interest can then be used for further analysis of, for example, catalytic activity, or as input for the GREG routine with a second reactant in order to probe competitive reactions.
\end{enumerate}

\section{\label{sec:computational}Computational Details}

The freeware Python3 program nanocut\cite{aradi_2019} is used to cut NPs from a ReaxFF-optimized FCC-Pt crystal structure ($a = 3.94737$~\AA). ReaxFF-GCMC calculations are performed using SCM’s implementation of ReaxFF\cite{scm_reaxff,vanduin2001,chenoweth2008} and the GCMC algorithm\cite{senftle2013,senftle2014} in the Amsterdam Density Functional Suite version 2017.106.\cite{scm_reaxff} The Pt-O force field developed in our group\cite{fantauzzi2014} is used. GCMC simulations are performed for each particle shape using oxygen chemical potentials $\mu_\text{O}$ corresponding to 200--1000~K at ultra-high vacuum (UHV, $p_\text{\ce{O2}} = 10^{-10}$~mbar) and corresponding to 400--1200~K at near-ambient pressure (NAP, $p_\text{\ce{O2}} = 1$~mbar) conditions. The latter choice is motivated by a recent combined experimental and theoretical study where NAP-XPS measurements on Pt(111) were able to confirm the presence of thermodynamically stable platinum surface oxide phases which were predicted using the GCMC scheme employed in this study.\cite{fantauzzi_growth_2017} The chemical potential of the gas phase oxygen reservoir, $\mu_\text{O}$, is obtained by correcting the ReaxFF total energy of \ce{O2} with empirical entropy and enthalpy contributions obtained from the NIST-JANAF thermochemical tables\cite{chase1998} and halving the obtained value. Convergence of the ReaxFF-GCMC runs is usually observed within 15,000--20,000 iterations in case of strongly oxidizing conditions (see Supporting Figure 1). Up to 1000 energy relaxation steps are performed between GCMC steps, using a convergence criterion of 0.5 kcal mol$^{-1}$. Full charge equilibration is performed using the EEM scheme.\cite{mortier_electronegativity_1985,mortier_electronegativity-equalization_1986,van_genechten_intrinsic_1987} 25,000 GCMC iterations are performed for each combination of temperature, pressure, and particle shape.

\begin{figure}[htbp]
	\includegraphics[width=0.67\linewidth]{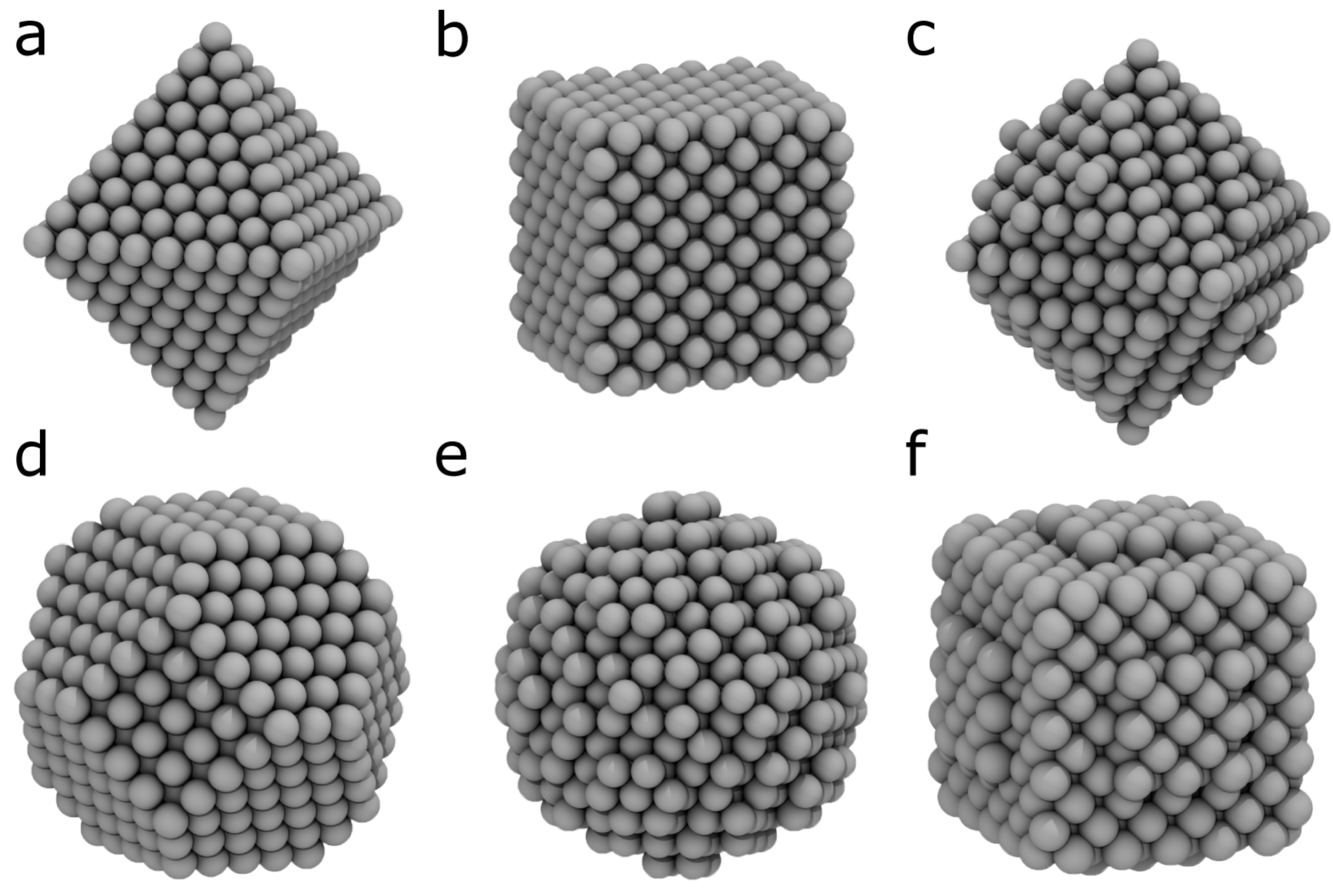}
	\caption{Illustrations of the 3 nm particle shapes used as model systems in this study. \textbf{a} Octahedron (\{111\}-faceted), \textbf{b} cube (\{100\}-faceted), \textbf{c} dodecahedron (\{110\}-faceted), \textbf{d} cuboctahedron (\{111\}- and \{100\}-faceted), \textbf{e} sphere (mixed-faceted), \textbf{f} tetrahexahedron (\{730\}-faceted).}
	\label{fgr:particles}
\end{figure}

The GREG routine is applied to the 3~nm NPs illustrated in Figure \ref{fgr:particles}.
This particle size is typical for industrial applications of Pt oxygen reduction electrocatalysts\cite{meier2014}. Since particles with defined shapes cannot be generated at arbitrary sizes, there is variance in the approximate diameter of the particles. Because particles have different surface-to-volume ratios, the number of atoms in each system varies, too. Table \ref{tbl:sizes} lists facet type, diameters, and number of atoms of the model systems in Figure \ref{fgr:particles}.
\begin{table}[htbp]
	\renewcommand{\arraystretch}{1.5}
	\caption{Structural features of the used shaped NPs.}
	\label{tbl:sizes}
	\begin{tabular}{cccc}
		\hline
		particle shape & facet type & \textit{d}$_{\text{approx.}}$ / nm & no. of atoms\\ \hline
		octahedron & \{111\} & 3.36 & 489 \\
		cube & \{100\} & 3.04 & 665  \\
		dodecahedron & \{110\} & 3.45 & 617 \\
		cuboctahedron & \{111\} + \{100\} & 3.21 & 711 \\
		sphere & mixed & 3.00 & 767 \\
		tetrahexahedron & \{730\} & 3.20 & 743 \\
		\hline
	\end{tabular}
\end{table}
The octahedral, cubic, and dodecahedral particle shapes were chosen as analogues to the experimentally well-studied Pt(111), Pt(100), and Pt(110) single crystal systems, as has been done in other computational studies.\cite{wen2009,wen2009_2,huang2011} The cuboctahedral shape was found to be the thermodynamically most stable shape in theoretical studies\cite{wen2009,wen2009_2,kirchhoff2019} and is compared here to the idealized spherical particle which also oftentimes serves as model for atomistic simulations of NPs.\cite{wen2009,wen2009_2} If no specialized, shape-specific synthetic procedure is invoked, experimental Pt NP catalysts in this size range are typically assumed to have (distorted) cuboctahedral shape.\cite{meier2014} The tetrahexahedral shape, first synthesized by Tian and co-workers,\cite{tian2007,tian2008} was considered in order to study the influence of high-indexed surface facets.

The cuboctahedral particle, which was the focus of a preceding study from our group,\cite{kirchhoff2019} will serve as a point of reference for comparing the effect of particle shape. The dataset for the cuboctahedral particle evaluated in this study was reproduced from scratch alongside the other particle shapes to ensure methodological consistency. No significant differences compared to the original dataset were found despite the stochastic nature of the ReaxFF-GCMC approach.

% -----------------------------------------------------------

\subsection{\label{sec:results}Results}

\subsubsection{GREG step 1: Model system and ReaxFF-GCMC calculations}

\begin{figure}[htbp]
	\centering
	\begin{minipage}[t][][b]{0.02\linewidth}
	    a
	\end{minipage}
	\begin{minipage}[t][][b]{0.46\linewidth}
        \includegraphics[width=\linewidth]{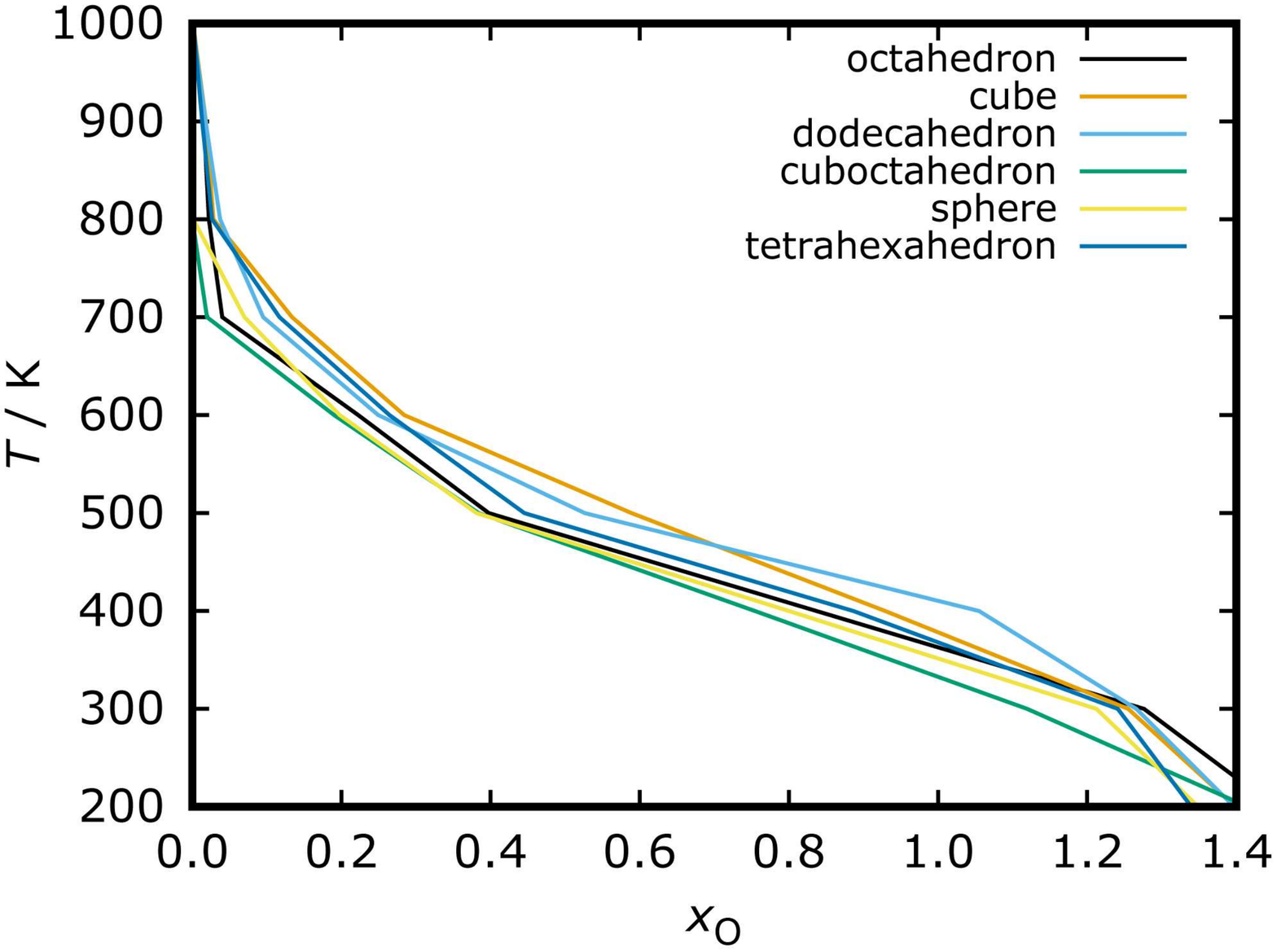}
	\end{minipage}
	\begin{minipage}[t][][b]{0.02\linewidth}
	    b
	\end{minipage}
	\begin{minipage}[t][][b]{0.46\linewidth}
        \includegraphics[width=\linewidth]{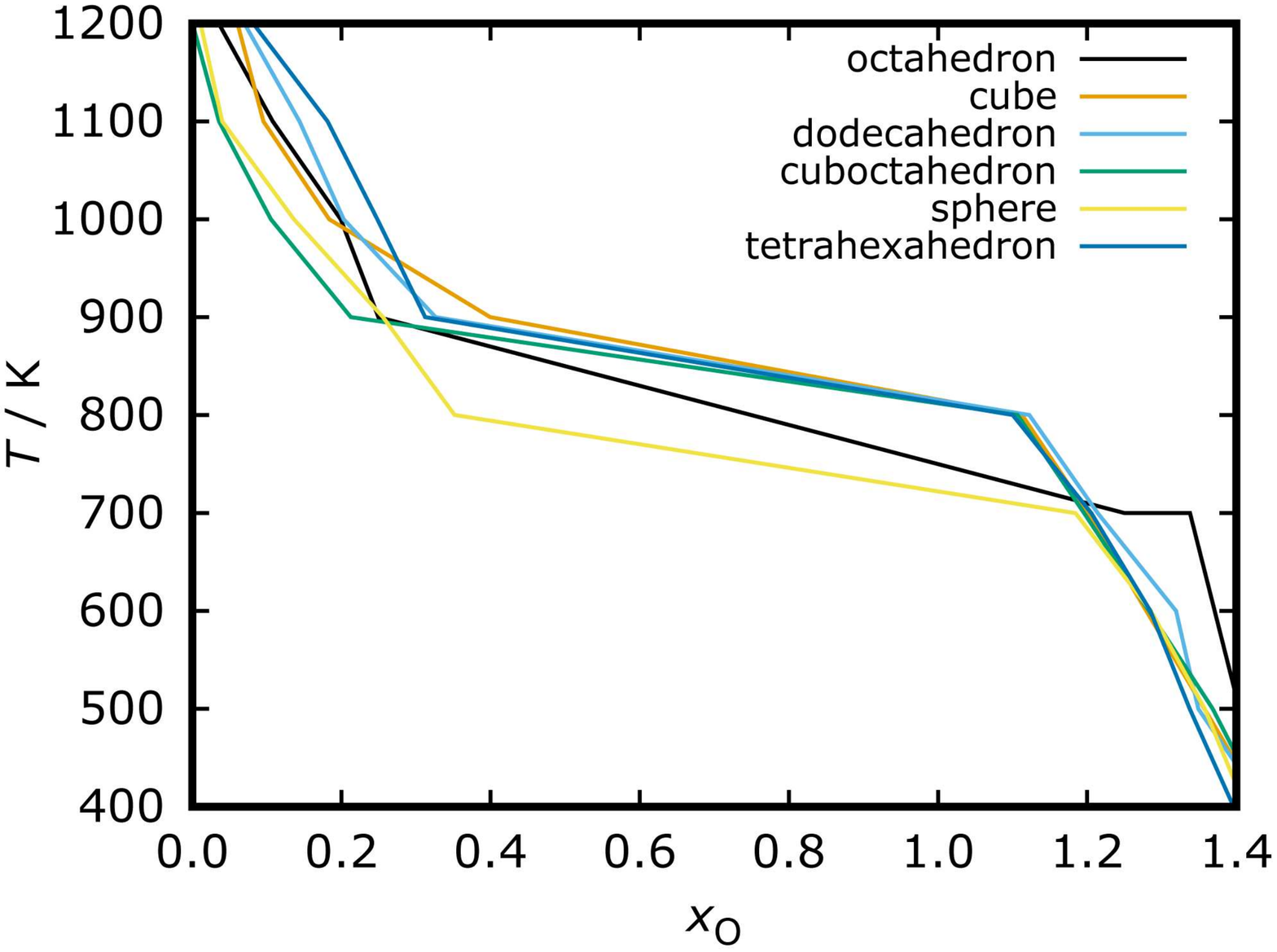}
	\end{minipage}  
	\caption{Oxygen adsorption isobars for the six tested particles (\{111\}-indexed octahedron, \{100\}-indexed cube, \{110\}-indexed dodecahedron, \{111\}- and \{100\}-indexed cuboctahedron, mixed-indexed sphere, and \{730\}-indexed tetrahexahedron) at (a) UHV and (b) NAP conditions. \textit{x}$_\text{O}$: oxygen-to-platinum ratio. \textit{x}$_\text{O}$ values are an average over the 5~\% lowest-energy structures at each temperature.}
	\label{fgr:isobars}
\end{figure}

Individual runs are analyzed with regards to the gas phase oxygen chemical potential, $\mu_\text{O}(T,p_\text{\ce{O2}})$. The oxygen-to-platinum ratio, $x_\text{O} = \frac{\text{\# of all O atoms}}{\text{\# of all Pt atoms}}$ will be used to quantify the degree of oxidation.\cite{senftle2014} By calculating $x_\text{O}$ of the 5~\% energetically most favorable structures generated at each temperature $T$ and showing the results as a function of $T$, adsorption isobars are generated, see Figure \ref{fgr:isobars}.

Adsorption isobars characterize the overall oxidation process, which can be divided into three stages. In the first stage from $x_\text{O} = 0.0$ to $x_\text{O} = 0.3$ (UHV: 1000--700~K, NAP: 1200--900~K), adsorption of oxygen atoms is observed mostly on the particles' vertices and edges. This initial stage will be referred to as surface oxidation. 
 
\begin{figure}[htbp]
	\includegraphics[width=0.67\linewidth]{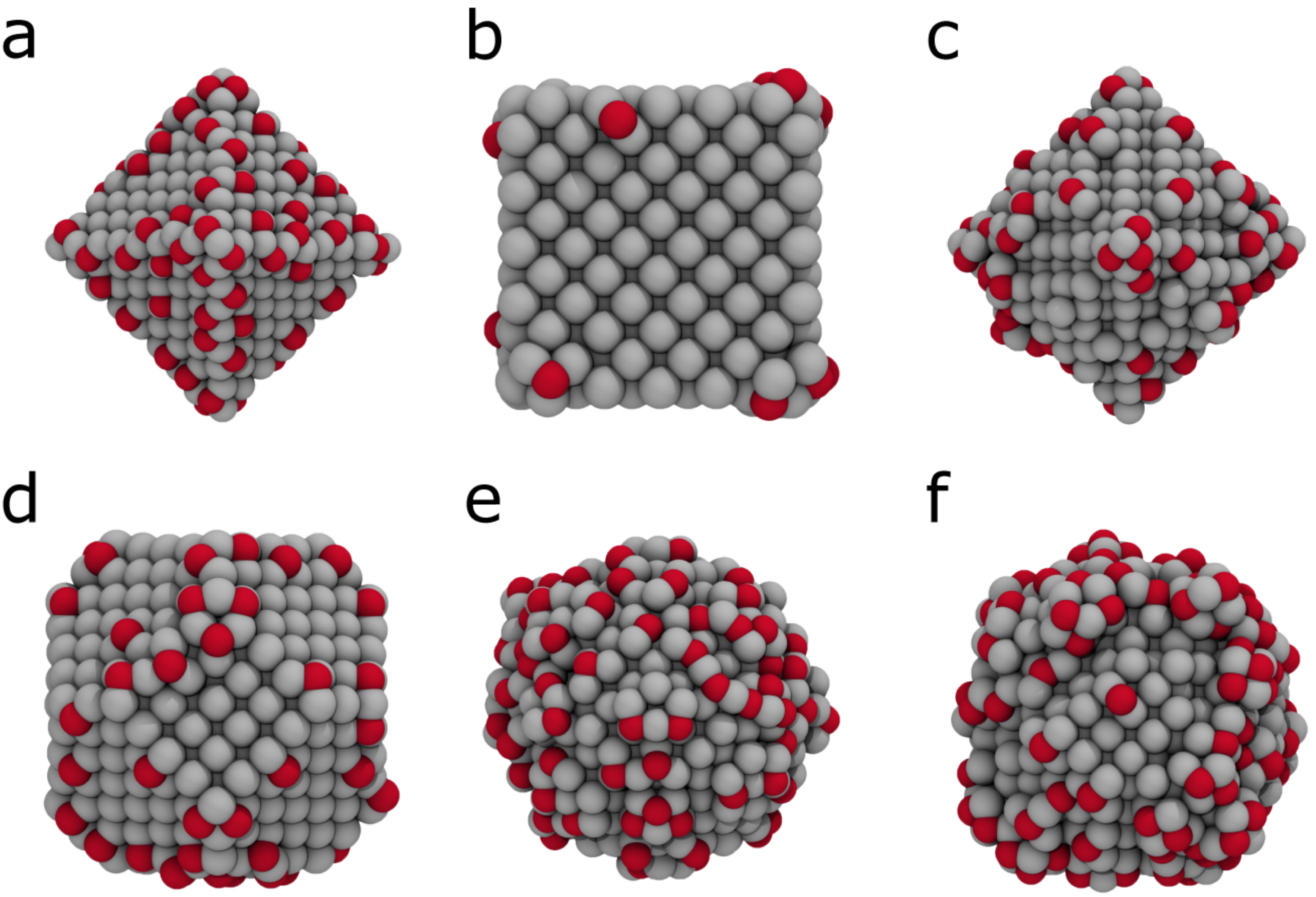}
	\caption{Exemplary illustrations of lightly oxidized particles generated under NAP conditions (see also Figure \ref{fgr:isobars} b). \textbf{a} Octahedron at 1000 K, \textbf{b} cube at 1100 K, \textbf{c} dodecahedron at 1100 K, \textbf{d} cuboctahedron at 1000 K, \textbf{e} sphere at 1000 K, and \textbf{f} tetrahexahedron at 1100 K.}
	\label{fgr:oxidationonset}
\end{figure}

While oxidation appears to proceed similarly for all particles from a macroscopic perspective of adsorption isobars, differences between particle shapes emerge under microscopic examination. Figure \ref{fgr:oxidationonset} shows exemplary structures of oxidized particles generated under NAP conditions between 1000--1100~K.
For the octahedron (Figure \ref{fgr:oxidationonset} \textbf{a}), cube (\textbf{b}), cuboctahedron (\textbf{d}), and to a lesser extent for the dodecahedron (\textbf{c}), the oxidation process starts at the vertices and edges of the particles. Facets are only lightly decorated. This trend is particularly apparent for the octahedron and cuboctahedron, where pristine facets are framed by oxidized edges. The sphere (\textbf{e}) and tetrahexahedron (\textbf{f}) show more homogeneous oxide coverage. For the tetrahexahedron in particular, significant changes to the particle surface structure are observed already at such low coverages of $x_\text{O} \approx 0.1$.

\begin{figure}[htbp]
	\includegraphics[width=0.67\linewidth]{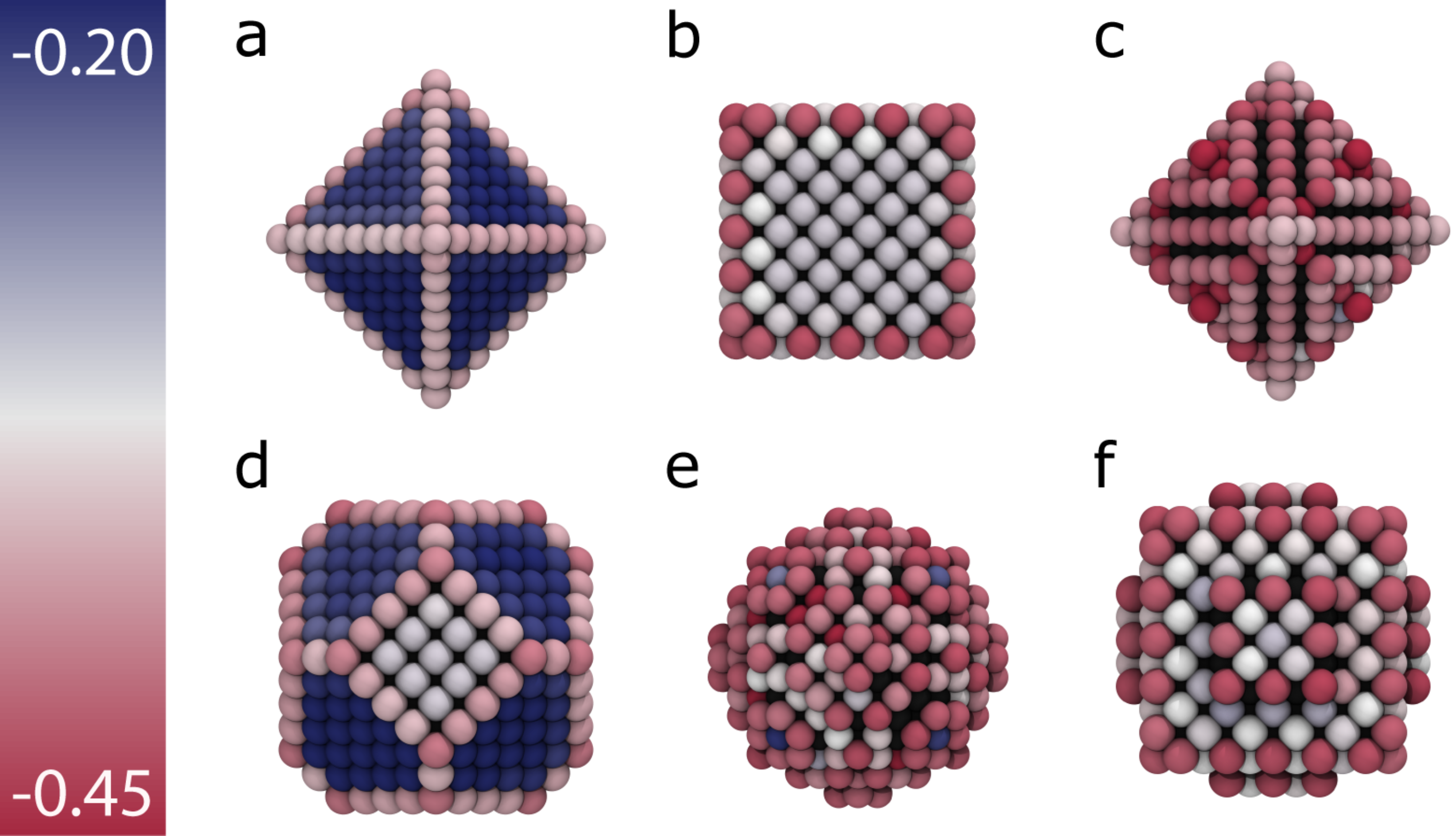}
	\caption{Heat maps illustrating the interaction strength between an \ce{O^2-} probe atom and the \textbf{a} octahedron, \textbf{b} cube, \textbf{c} dodecahedron, \textbf{d} cuboctahedron, \textbf{e} sphere, and \textbf{f} tetrahexahedron. Interaction strength increases in the sequence from blue to white to red. Scale is  given in eV.}
	\label{fgr:heatmaps}
\end{figure}

The different early oxidation trends can be rationalized in the context of interaction energy heat maps, see Figure \ref{fgr:heatmaps}.
The heat maps illustrate thermodynamic differences in oxygen affinity between particle shapes. Interaction strength between the \ce{O^2-} probe and the surface facets is declining in the sequence \{110\} $>$ \{100\} $>$ \{111\}. As a result, particles with \{111\} or \{100\} facets --- \textit{i.e.}, octahedron, cube, and cuboctahedron --- are initially oxidized mostly at the more reactive edges and vertices (see Figure \ref{fgr:oxidationonset}). The more open \{110\} facets of the dodecahedron show similar interaction strength to its edges and vertices, resulting in homogeneous, less ordered oxidation similar to the mixed-faceted sphere and the \{730\}-indexed tetrahexahedron.

\begin{figure}[htbp]
	\includegraphics[width=0.67\linewidth]{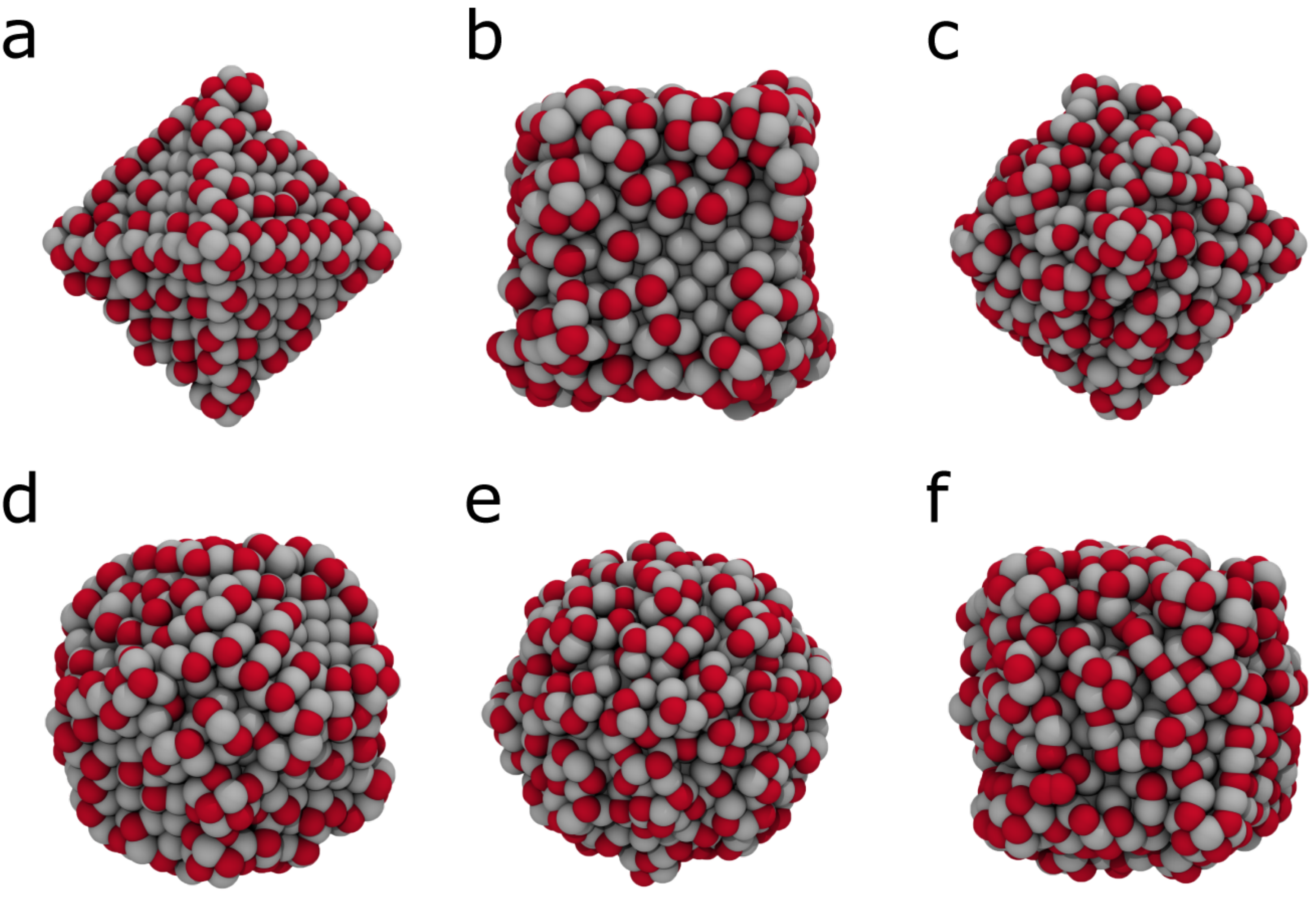}
	\caption{Exemplary illustrations of more strongly surface-oxidized particles ($x_{\text{O}} \approx 0.3$). \textbf{a} Octahedron at 850 K, \textbf{b} cube at 850 K, \textbf{c} dodecahedron at 850 K, \textbf{d} cuboctahedron at 850 K, \textbf{e} sphere at 850 K, and \textbf{f} tetrahexahedron at 800 K.}
	\label{fgr:surfaceoxides}
\end{figure}

The differences in oxidation trends become more pronounced as the surface oxide coverage increases to $x_\text{O} \approx 0.3$, see Figure \ref{fgr:surfaceoxides}.
Both the octahedron (\textbf{a}) and the cuboctahedron (\textbf{b}) largely retain their overall shape. Notably, the \{111\}-indexed facets remain comparatively free of adatoms. Analoguous to the cubic particle (\textbf{b}), the  \{100\}-indexed facets of the cuboctahedron are more strongly oxidized than the \{111\} facets and oxygen atoms can already be found in the subsurface. While \{100\}- and \{111\}-indexed facets are similar in behavior at low coverages (Figure \ref{fgr:oxidationonset}), the difference in interaction strength with oxygen adatoms illustrated in the heat maps (Figure \ref{fgr:heatmaps}) at this stage of oxidation leads to a higher density of adatoms on \{100\} facets. The dodecahdral (\textbf{c}), spherical (\textbf{e}), and tetrahexahedral (\textbf{f}) particles show homogeneous oxygen coverage. Both the dodecahedral and tetrahexahedral particles appear significantly more rounded than at $x_\text{O} = 0.1$ and their initial facet structure is lost.

The second stage of oxidation is characterized by a sudden increase of $x_\text{O}$ from \textit{ca.}~0.3 to 1.2 at a critical temperature of \textit{ca.}~850~$\pm$~50~K under NAP conditions. At UHV oxygen pressure, the increase is more gradual. Beyond $x_\text{O} \approx 1.2$, structures are fully oxidized and the dominant structural motif is of the previously reported agglomeration of clusters of \ce{Pt6O8} stoichiometry.\cite{kirchhoff2019} This stage likely indicates a point of full degradation of the particle since its original shape and corresponding reactivity-defining surface features are lost and cannot be recovered by reductive measures anymore. The \ce{Pt6O8} clusters consist of a Pt cube with O atoms located on the sides of the cube. DFT and ReaxFF calculations showed that formation of individual \ce{Pt6O8} clusters from the elements is thermodynamically favorable, and that detachment of a \ce{Pt6O8} cluster from a fully oxidized structure is strongly exothermic if water is present during the process to stabilize the cluster.\cite{kirchhoff2019} This stage will be referred to as the fully oxidized phase. The resulting particles are amorphous - independent of initial facet structure - as indicated by Pt-Pt radial distribution functions (see Supporting Figure 2). Partial charge distribution analysis using the EEM method suggests that this [\ce{Pt6O8}]$_n$ structure is most closely related to the mixed Pt(II) / Pt(IV) \ce{Pt3O4} bulk oxide.\cite{kirchhoff2019}

The last stage of oxidation starts at $x_\text{O} \approx$~1.2--1.3 and is characterized by decoration of the oxidized particles with dioxygen species. This type of structure is only observed under harsh conditions (NAP, 400--700~K); at UHV conditions (Figure \ref{fgr:isobars}\textbf{a}) this species is not observed.

\subsubsection{GREG step 2: Merging of datasets and construction of potential-dependent phase diagrams}

\begin{figure*}[htbp]
	\includegraphics[width=\linewidth]{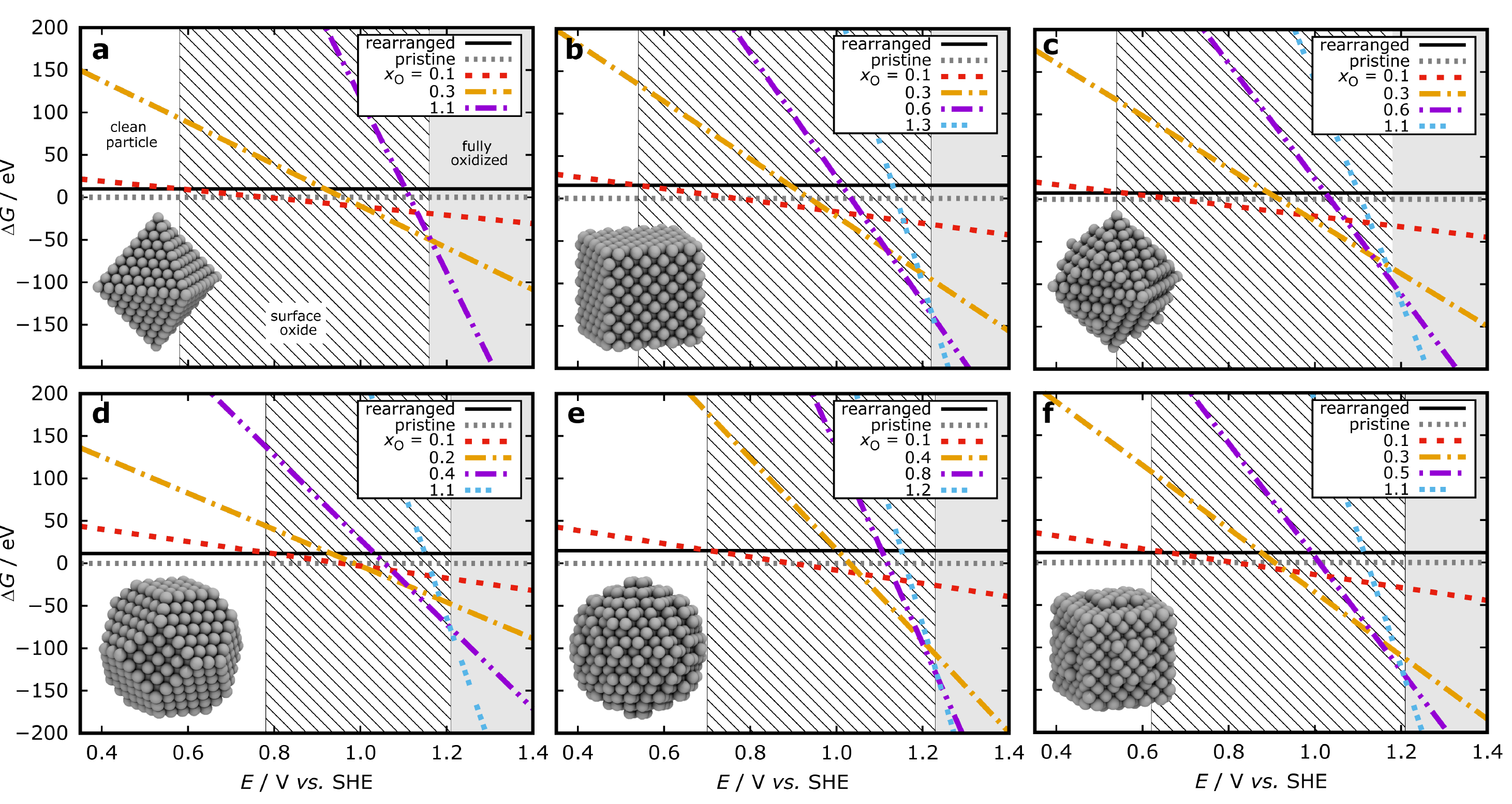}
	\caption{Electrochemical phase diagrams generated using data from the GCMC procedure and the EAITD approach for 3 nm \textbf{a} octahedral, \textbf{b} cubic, \textbf{c} dodecahedral, \textbf{d} cuboctahedral,\cite{kirchhoff2019} \textbf{e} spherical, and \textbf{f} tetrahexahedral particles. Types of shading indicate different phases: no shading, clean particle; diagonal stripes, adsorbate and surface oxide up to $x_\text{O} = 0.3$; solid filling, fully oxidized structure. Diagrams have been filtered to only display the lowest-energy structures to maintain easy readability. Results for the cuboctahedral particle (d) are reproduced from Kirchhoff \textit{et al.}\cite{kirchhoff2019}}
	\label{fgr:phasediagrams}
\end{figure*}

Electrochemical phase diagrams are constructed by merging all datasets generated by the GCMC algorithm for the same particle shape and finding the most stable structures as a function of the electrode potential using EAITD theory.\cite{jacob2007} All GCMC output structures for each particle shape are grouped based on their $x_\text{O}$ with a bin width of $\Delta x_\text{O} = 0.1$. The most stable structure from each bin according to eq. \eqref{eq:potentialdependent} are represented as the intersecting lines in Figure \ref{fgr:phasediagrams}.
Only the enveloping, most stable structures are shown for clarity. Analogous to a preceding study from our group which focused on oxidation of cuboctahedral particles exclusively,\cite{kirchhoff2019} both pristine NPs and slightly rearranged particles are given as reference systems. The rearranged reference particles are generated by stripping all oxygen atoms from surface-oxidized structures and relaxing them, allowing for rearrangement. These structures are given as a secondary reference that is likely more representative of the \textit{in-situ} structure during continuous oxidation and reduction cycles compared to the perfectly crystalline starting structures. The reference energy values of the rearranged particles are an average over 5 structures obtained via the procedure outlined above, using different oxidized structures. Based on the total energy fluctuations of the averaged rearranged structures, an uncertainty of the surface oxidation onset potentials of 0.05--0.10~V is estimated (see Supporting Information for more details), which is in a similar order of magnitude as the general uncertainty expected from the ReaxFF model. In the following, three phases are considered: the clean particle phase extends up to the intersection between the 'rearranged' phase and the first phase representing surface oxidation ($x_\text{O}$~=~0.1--0.3); the adsorption and surface oxide phase, where oxygen atoms are incorporated into the surface layers, extends between this point and the intersection with a fully oxidized structure ($x_\text{O} \approx 1.2$); the fully oxidized phase featuring the \ce{Pt6O8} motif extends beyond this point.

\subsubsection{GREG step 3: Analysis of electrochemically most stable structures}

Analyzing the  electrochemical phase diagrams in Figure \ref{fgr:phasediagrams} in more detail, a first observation concerns the onset potentials for surface oxidation as a function of NP shape, see comparison in Table \ref{tbl:onsetpotentials}.
\begin{table}[htbp]
	\renewcommand{\arraystretch}{1.5}
	\caption{Onset of surface and full particle oxidation in V \textit{vs.} SHE for the 3 nm particles.}
	\label{tbl:onsetpotentials}
	\begin{tabular}{cccc}
		\hline
		particle shape & surface oxidation & surface oxidation & full oxidation \\
		& (rearranged reference) & (pristine reference) & \\\hline
		octahedron & 0.58 V & 0.79 V & 1.16 V \\
		cube & 0.54 V & 0.76 V & 1.22 V  \\
		dodecahedron & 0.54 V & 0.68 V & 1.19 V \\
		cuboctahedron\cite{kirchhoff2019} & 0.78 V & 0.94 V & 1.21 V \\
		& 0.84 V$^\dagger$ & --- & 1.15 V$^\dagger$ \\
		sphere & 0.70 V & 0.89 V & 1.23 V \\
		tetrahexahedron & 0.62 V & 0.81 V & 1.21 V \\
		\hline
		\multicolumn{4}{l}{\small$^\dagger$ Including solvation and entropy contributions  obtained by Kirchhoff \textit{et al}.\cite{kirchhoff2019}}
	\end{tabular}
\end{table}
The octahedron, the cube, and the dodecahedron start to oxidize first, with onset potentials as low as 0.5--0.6~V with respect to the rearranged clean reference. The tetrahexahedron follows with an onset potential slightly above 0.6~V. The sphere and cuboctahedron require much harsher conditions, with onset potentials of \textit{ca.}~0.7 and 0.8~V, respectively. Note that the electrochemical phase diagrams presented here do not include solvation or thermochemical contributions. These contributions were previously shown to add up to an increase of the onset potential for surface oxidation and a decrease of the onset potential for complete oxidation by 0.05--0.10~V.\cite{kirchhoff2019} If pristine, non-rearranged NPs are considered as reference, onset potentials for surface oxidation are systematically shifted to more positive potentials by \textit{ca.}~0.2~V.

\begin{figure}[htbp]
    \includegraphics[width=0.67\linewidth]{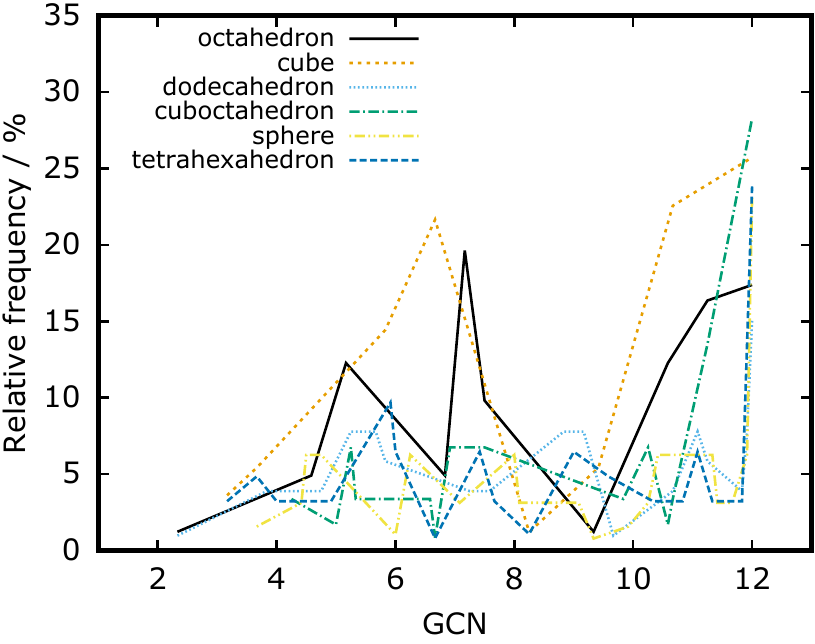}
	\caption{Frequency of occurrence of each generalized coordination number ($GCN$), given as relative percentage of the total number of atoms in the NP.}
	\label{fgr:gcn_charge_a}
\end{figure}

The surface oxidation onset potential trends can be rationalized by considering geometric and electronic effects. Firstly, a notable difference between systems \textbf{a--c} and \textbf{d--f} in Figure \ref{fgr:phasediagrams} is that the latter three particles have a lower surface-to-volume ratio (see also Table \ref{tbl:sizes}). For example, the cuboctahedral particle is an octahedron with flattened vertices and the tetrahexahedron is a cube with rounded edges and less open facet structure; both the cuboctahedron and the tetrahexahedron therefore have a smaller number of atoms with low coordination numbers at their edges and vertices. These differences can be quantified by analyzing the particles in terms of generalized coordination numbers ($GCN$s)\cite{calle2014} of the Pt atoms. The $GCN$ of an atom $i$ is defined as
\begin{equation}
GCN(i) = \sum_{j=1}^{n_j} \frac{CN(j)}{CN_{\text{max}}}\, , \label{gcns}
\end{equation}
where $CN(j)$ are the classical coordination numbers of neighboring atoms $j$ and $CN_{\text{max}}$ is a weighing factor. $CN_{\text{max}}$ gives the largest possible coordination number and is 12 for an FCC crystal, 9 for a BCC crystal, \textit{etc}.
Figure \ref{fgr:gcn_charge_a} illustrates the frequency of $GCN$ values for each particle shape.
Lower $GCN$ values correspond to more dangling bonds of that particular atom whereas GCNs increase for atoms closer to the particle center. The maximum $GCN$ value is 12 for the NPs studied here.

Particles with a low surface-to-volume ratio such as the sphere, cuboctahedron, and tetrahexahedron show an overall homogeneous distribution of $GCN$s. On the opposite, the octahedron and cube show a notably higher percentage of smaller $GCN$s than the other systems. The abundance of undercoordinated atoms therefore correlates with more negative surface oxidation onset potentials for the octahedron and cube. However, the early oxidation onset of the dodecahedron, for which an overall homogeneous $GCN$ distribution is obtained as well, cannot be fully explained by this geometric descriptor.

\begin{figure}[htbp]
    \includegraphics[width=0.67\linewidth]{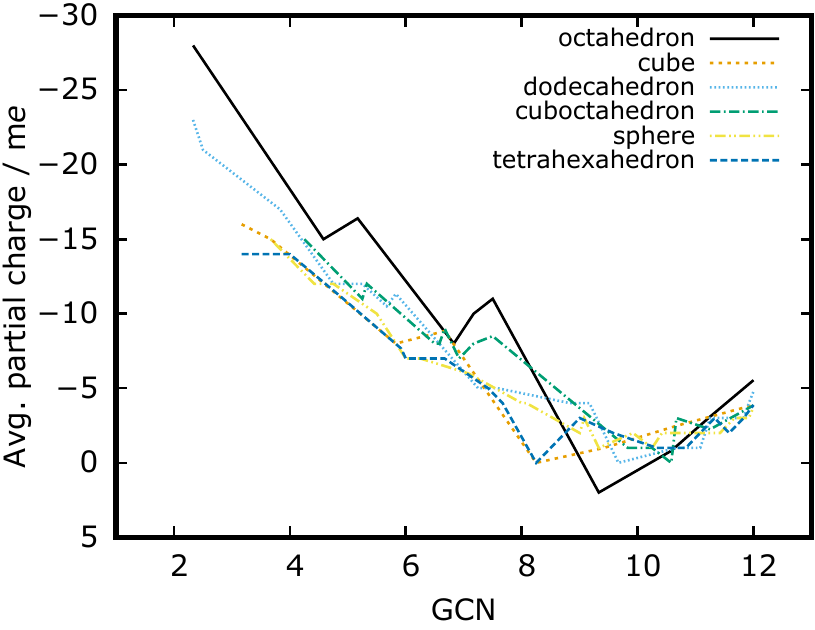}
	\caption{Average partial charge calculated for atoms of a specific $GCN$.}
	\label{fgr:gcn_charge_b}
\end{figure}

\begin{figure}[htbp]
	\includegraphics[width=0.67\linewidth]{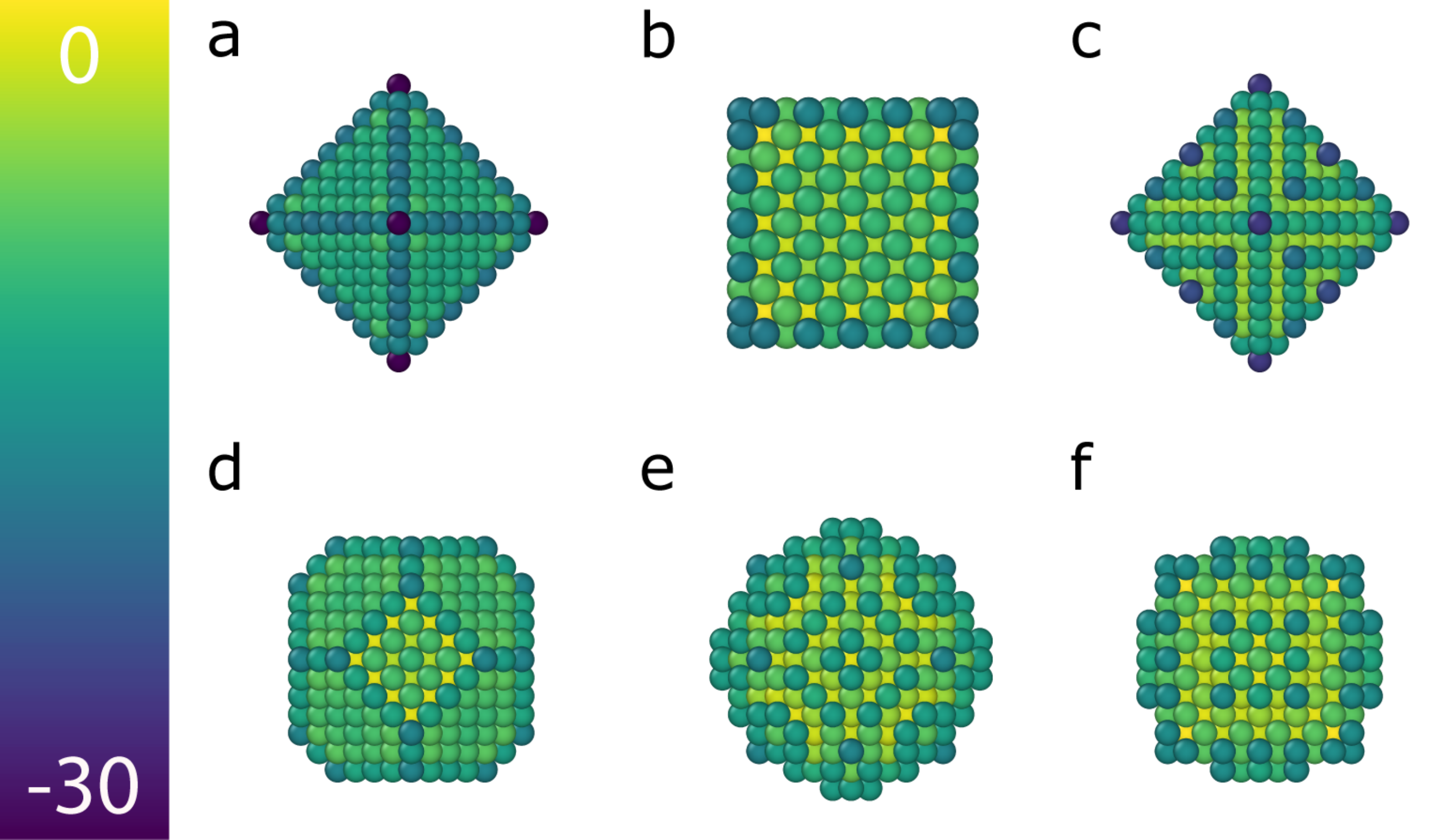}
	\caption{Illustration of average partial charges on \textbf{a} octahedron, \textbf{b} cube, \textbf{c} dodecahedron, \textbf{d} cuboctahedron, \textbf{e} sphere, and \textbf{f} tetrahexahedron. Particles carry a total of 4 negative charges. Scale is given in m\textit{e}.}
	\label{fgr:charges}
\end{figure}

To investigate electronic effects, partial charge distributions are invoked as an electronic descriptor. 4 \ce{e-} of test charges --- assuming the 4 \ce{e-} ORR pathway as benchmark case --- are introduced to the systems and redistribution of the charges across the particles is studied using the EEM charge equilibration scheme. Figure \ref{fgr:gcn_charge_b} shows the average partial charge on atoms as a function of their $GCN$ and Figure \ref{fgr:charges} illustrates the distribution of partial charges on the particles.
This analysis reveals that atoms with low $GCN$s carry a comparatively large charge in case of the octahedron and, notably, the dodecahedron. Figure \ref{fgr:gcn_charge_b} therefore reveals that accumulation of charges on the edges and vertices of the dodecahedron is a likely cause for its early oxidation onset in the corresponding phase diagrams.

\subsection{\label{sec:discussion}Discussion}

The GREG routine was used to simulate oxidation of 3~nm NPs of various shapes. From the simulation results, trends connecting particle shape and oxidative stability can be derived and will be discussed in the following.

Under non-electrochemical oxidation conditions, a broad ranking of the different particle shapes can be made with regards to their proclivity towards oxidation. Within this first phase of (surface) oxidation ($x_\text{O}$~=~0.0--0.3), the cuboctahedral and spherical particles appear to be most resistant towards oxidation as lower $T$ are required to achieve the same $x_\text{O}$ compared to other particles; see Figure \ref{fgr:isobars}. The tetrahexahedral, cubic, and dodecahedral particles, on the other hand, appear to oxidize more readily. The octahedral particle is found in between those extremes. Notably, there are differences in the rankings of particle shapes between NAP and UHV conditions. This result could indicate that higher oxidative stress could change the oxidation behavior. However, it cannot be ruled out that these difference are in part due to the stochastic nature of the GCMC algorithm. Furthermore, note that the different (pristine) particles show different thermodynamic stability.\cite{kirchhoff2019} Therefore, this simulation approach cannot account for the possibility of a lower-stability particle shape rearranging into a more stable shape at ambient conditions, either by means of temperature or by surface-adsorption induced rearrangement.\cite{zhu2013}

To the best of the authors' knowledge, only one other group studied the oxidation Pt NPs using the ReaxFF-GCMC methodology. Gai \textit{et al.} combined ReaxFF-GCMC with reactive molecular dynamics to study O and H uptake in a stepped Pt(321) surface and octahedral, cuboctahedral, and cubic Pt NPs.\cite{gai2016} The group calculated adsorption isotherms and made general observations about the relationship between the number of available edge and vertex sites for each NP shape and the resulting proclivity towards oxidation, which the present results are in agreement with. However, for NP oxidation, our simulations suggest that \{111\}-indexed facets are more resistant towards oxidation than \{100\} facets whereas Gai \textit{et al.} report the opposite trend. Oxygen adsorption energy results from DFT calculations on Pt(111) and Pt(100) single crystal surfaces indicate that adsorption on Pt(100) is preferred over Pt(111) at high coverage but values reported for low coverage are very similar.\cite{gu_absorption_2007} Such small differences in adsorption energy strain the limits of the ReaxFF method. Thus, the disagreement between the present work and the study by Gai \textit{et al.} with regards to oxidative stability of the \{111\} and \{100\} facets is likely insignificant.

For $x_\text{O}$~>~\textit{ca.}~1.2, particles are found to become amorphous and decompose into clusters of \ce{Pt6O8} stoichiometry. This stage of oxidation is likely a 'point of no return' after which the original shape of the particles cannot be regenerated anymore, for example by reductive potential jumps in electrochemical applications. Oxidation beyond this point should be avoided to ensure longevity of the catalyst. Notably, Gai \textit{et al.} show an incomplete structure reminiscent of the \ce{Pt6O8} motif in their study which formed on a Pt(321) model surface.\cite{gai2016} However, they do not discuss this result any further and did not continue simulations until systems were fully oxidized.

To fully explain the sequence of electrochemical onset potentials for surface oxidation of the different particle shapes obtained from electrochemical phase diagrams (Figure \ref{fgr:phasediagrams}), both $GCN$ and partial charge distribution analysis were carried out. This observation highlights the importance of taking both geometrical and electronic effects into account when studying reactivity trends in catalysis.\cite{braunwarth2020} Beyond the clean particle, surface oxide, and fully oxidized (\ce{[Pt6O8]_$n$}) phases in the electrochemical phase diagrams, a fourth phase representing the bulk oxide, $\alpha$-\ce{PtO2}, would technically be expected because \ce{$\alpha$-PtO2} (\textit{i.e.} fully oxidized Pt(IV)) is the thermodynamic end point for the oxidation of Pt. However, given the molecular nature of the used model systems, rearrangement into the sheet-like $\alpha$-\ce{PtO2} structure is unfavorable and thus, the simulation does not proceed beyond the \ce{[Pt6O8]_$n$} phase. According to experimental results and previous DFT studies using the same approach, $\alpha$-\ce{PtO2} should become stable beyond 1.2--1.3~V \textit{vs.} SHE.\cite{jacob2007} This would mean that NPs consisting of \ce{Pt6O8} units solely should have a small stability region - depending on the cluster shape - between \textit{ca}. 1.1 and 1.3~V.

Only few experimental reports of measured oxidation onset potentials are available for comparison with the present simulations. Merte \textit{et al.} report onset of O surface adsorption for small (\textit{ca.} 1.2~nm) size- but not shape-selected NPs at \textit{ca.}~0.3~V~\textit{vs.}~RHE and formation of stable oxides beyond \textit{ca.}~1.0~V.\cite{Merte2012} While these values are in good agreement with the present study when taking into account the size-dependent shift of surface and complete oxidation onset voltages reported in our previous work on 2--4~nm cuboctahedral particles,\cite{kirchhoff2019} there are, to the best of our knowledge, no experimental measurements of electrochemical oxidation of both size- and shape-selected NPs that would allow for rigorous comparison. Substrate effects complicate comparability with experiments even further. For example, Ono \textit{et al.} report that flattening of oxidized Pt NPs is observed on substrates and that the degree of flattening increases with increasing NP/support interaction strength.\cite{Ono2010} Future studies on Pt NP oxidation should therefore aim to include the substrate explicitly to further close the gap between model and experiment.

Finally, note that results from these oxidation simulations cannot be directly linked to experimentally measured ORR activity. The reaction mechanism of the ORR has not been considered in full; the present work focuses only on the (albeit critical) O intermediate and oxidative degradation in particular. Note however that experimental studies comparing the ORR activity of various NP shapes have so far been inconclusive. For example, Wang \textit{et al.} compared the ORR activity of 3--7~nm Pt polyhedra, truncated cubes, and cubic particles.\cite{wang2008} Their measurements showed the cube to outperform all other tested materials. Another study by Narayanan \textit{et al.} reports ORR activities for 5--7~nm tetrahedral, cubic, and near-spherical nanoparticles but found tetrahedral NPs to be most active.\cite{narayanan2004} Future investigations should therefore focus on using the oxidized NPs obtained in the present work to study adsorption and transformation of the other ORR intermediates to give reactivity estimates and guide experimental ORR activity measurements.

% -----------------------------------------------------------

\section{\label{sec:conclusions}Conclusions}

A simulation procedure for electrochemical reactions on nanoparticulate catalysts was presented. The procedure is referred to as GREG; the acronym highlights the important ingredients of this routine, namely usage of a \textit{grand-canonical} Monte Carlo algorithm in combination with \textit{reactive} force fields to study \textit{electrochemical} reactions as a function of the \textit{galvanic} cell potential.

Using the GREG procedure, the electrochemical oxidative degradation of 3 nm octahedral, cubic, dodecehedral, cuboctahedral, spherical, and tetrahexahedral nanoparticles was studied. The oxidized structures were analyzed in the context of adsorption isobars and the overall oxidation process can be divided into three stages: 1) surface oxidation, starting at edges and vertices; 2) full oxidation involving disassembly of the original particle shape into an agglomerate of \ce{Pt6O8} clusters; 3) decoration of the particle surface with dioxygen species. 

Electrochemical phase diagrams constructed from the oxidized structures reveal that surface oxidation occurs at lower electrode potentials for the octahedral, cubic, and dodecahedral particles (0.5--0.6~V \textit{vs.} SHE) than for the tetrahexahedral (0.6~V), spherical (0.7~V), and cuboctahedral (0.8~V) particles. These differences were rationalized using structural (generalized coordination numbers) and electronic (partial charge distributions, interaction energy heat maps) analysis. Complete oxidation is observed beyond \textit{ca.}~1.1--1.2~V \textit{vs.} SHE in all cases. 

The GREG procedure proves to be a powerful tool to probe stability trends for model catalysts in electrochemical reactions and thus enables understanding-driven improvement of electrocatalysts.

\begin{acknowledgement}

The authors gratefully acknowledge financial support by the Deutsche Forschungsgemeinschaft (DFG) through the collaborative research center SFB-1316 as well as the priority program SPP-2080. The authors also acknowledge support by the state of Baden-Württemberg through bwHCP and DFT through grant no INST 37/935-1 FUGG. The Volkswagen Group Wolfsburg is acknowledged for partial funding, as well as the Icelandic Research Fund. BK acknowledges the University of Iceland Research Fund for funding through a PhD fellowship.

CRediT author statement: \textbf{Björn Kirchhoff}: investigation, formal analysis, data curation, validation, writing - original draft; \textbf{Christoph Jung}: investigation, writing - review and editing; \textbf{Hannes Jónsson}: resources, supervision, writing - review and editing; \textbf{Donato Fantauzzi}: conceptualization, supervision; \textbf{Timo Jacob}: project administration, conceptualization, funding acquisition, resources, writing - review and editing, supervision.

\end{acknowledgement}

\begin{suppinfo}

Exemplary convergence behavior of the GCMC simulations and simulated XRD patterns of a pristine and a strongly oxidized cuboctahedral NP are shown in the Supporting Information.

The dataset of oxidized structures generated by ReaxFF-GCMC simulations is available via \url{https://doi.org/10.5281/zenodo.6322004}.

\end{suppinfo}

\bibliography{gcmc}

\end{document}